\documentclass[twocolumn]{aastex62}
\usepackage{natbib}
\usepackage{amsmath}
\usepackage{mathtools}

\received{March 18, 2020}
\revised{May 27, 2020}
\accepted{June 9, 2020}

\shorttitle{Deep Learning Prediction of Ly$\alpha$ emission}
\shortauthors{Hassan Fathivavsari}

\begin{document}

\title{\bf{Deep Learning Prediction of Quasars Broad Ly$\alpha$ Emission Line}}

\correspondingauthor{Hassan Fathivavsari}
\email{h.fathie@gmail.com}

\author{Hassan Fathivavsari}
\affiliation{School of Astronomy, Institute for Research in Fundamental Sciences (IPM), P. O. Box 19395-5531, Tehran, Iran}

\begin{abstract}
We have employed deep neural network, or \textit{deep learning} to predict the flux and the shape of the broad Ly$\alpha$ emission lines in the spectra of quasars. We use 17\,870 high signal-to-noise ratio (SNR\,$>$\,15) quasar spectra from the Sloan Digital Sky Survey (SDSS) Data Release\,14 (DR14) to train the model and evaluate its performance. The Si\,{\sc iv}, C\,{\sc iv}, and C\,{\sc iii]} broad emission lines are used as the input to the neural network, and the model returns the predicted Ly$\alpha$ emission line as the output. We found that our neural network model predicts quasars continua around the Ly$\alpha$ spectral region with $\sim$\,6$-$12\,\% precision and $\lesssim$\,1\% bias. Our model can be used to estimate the H\,{\sc i} column density of eclipsing and ghostly damped Ly$\alpha$ (DLA) absorbers as the presence of the DLA absorption in these systems strongly contaminates the flux and the shape of the quasar continuum around Ly$\alpha$ spectral region. The model could also be used to study the state of the intergalactic medium during the epoch of reionization.

\end{abstract}


\section{Introduction} \label{sec:intro}

\defcitealias{2018MNRAS.477.5625F}{Paper\,I}

The damped Ly$\alpha$ (DLA) absorption detected in the spectra of distant quasars are signatures of the presence of neutral clouds with H\,{\sc i} column density log\,$N$(H\,{\sc i})\,$\ge$\,20.3 along the line of sight \citep{1972ApJ...171..233L, 1986ApJS...61..249W, 1989ApJ...344..567T, 2005pgqa.conf..125R, 2005ARA&A..43..861W, 2008ApJ...681..881W}. These so-called DLA absorbers are observed at most redshifts \citep{2009MNRAS.394L..61K, 2011ApJ...732...35M, 2019ApJ...885...59B}. If the velocity separation between a DLA and its background quasar is $\Delta\,V$\,$\lesssim$\,1500\,km\,s$^{-1}$, i.e. the DLA and the quasar are almost at the same redshift, the DLA could act as a natural coronagraph blocking most of the Ly$\alpha$ emission emanating from the quasar's broad line region (BLR). When this happens, depending on the dimension of these so-called \emph{eclipsing} DLAs, weaker Ly$\alpha$ emission from the narrow line region of the quasar and/or star-forming regions in the quasar host galaxy could be detected as some narrow Ly$\alpha$ emission in the DLA absorption trough \citep{2009ApJ...693L..49H, 2013A&A...558A.111F, 2015MNRAS.454..876F, 2016MNRAS.461.1816F, 2018MNRAS.477.5625F, 2020ApJ...889L..12D}. If the size of an eclipsing DLA is much smaller than the BLR, the Ly$\alpha$ emission from the regions of the BLR that are not covered by the DLA, could fill the DLA absorption trough, and form a \emph{ghostly} DLA \citep{2017MNRAS.466L..58F, 2020ApJ...888...85F}. Such absorbers are called ghostly DLAs as no DLA absorption is seen in their spectra. 

Although the DLA absorption is detected in the spectrum of a quasar with an eclipsing DLA, it is still not straightforward to determine the DLA H\,{\sc i} column density. Since the DLA absorption falls on top of the quasar broad Ly$\alpha$ emission, one would need to know the \emph{intrinsic} (i.e. unabsorbed) shape of the quasar Ly$\alpha$ emission in order to be able to properly reconstruct the DLA damping wings, and accurately determine the H\,{\sc i} column density \citep{1999MNRAS.303..711L, 2019ApJ...885...59B}. In the case of ghostly DLAs, where no Ly$\alpha$ absorption is present in the spectrum, knowledge of the quasars Ly$\alpha$ emission could still allow us to estimate the H\,{\sc i} column density. Moreover, if the luminosity of the quasar broad Ly$\alpha$ emission is known, it could also help shed some light on the origin of the narrow Ly$\alpha$ emission detected in the eclipsing DLAs absorption trough \citep{2006A&A...459..717C, 2008A&A...488...91C, 2012A&A...542A..91N, 2016MNRAS.461.1816F}. 

Another important application of the knowledge of the quasars intrinsic Ly$\alpha$ emission is the study of the epoch of reionization. Proper reconstruction of the damping wing of the Ly$\alpha$ absorption in the spectra of high redshift ($z_{\rm QSO}$\,$\gtrsim$\,7) quasars provide important information on the state of the intergalactic medium during reionization \citep{2008ApJ...672...48C,2011MNRAS.416L..70B,2015MNRAS.454..681K,2017MNRAS.466.4239G,2018ApJ...864..142D,2020MNRAS.493.4256A}.

Over the years, various empirical approaches have been introduced to predict quasars continua around Ly$\alpha$ spectral region: 1) applying Gaussian fits to the red side of the Ly$\alpha$ emission line \citep{2009MNRAS.400.1493K}, 2) extrapolating quasars continua using properties of the emission lines located to the red side of the Ly$\alpha$ emission \citep{2012Natur.492...79S, 2018Natur.553..473B}, 3) exploiting the co-variance matrix of emission line properties \citep{2017MNRAS.466.1814G}, and 4) using principle component analysis (PCA) decomposition \citep[e.g.][]{2006ApJS..163..110S, 2018ApJ...864..143D}. These techniques are mainly based on relating the correlated spectral properties of the broad emission lines. 

\citet{2018ApJ...864..143D} have recently applied PCA decomposition technique on a sample of 12\,764 quasar spectra from SDSS/BOSS, and construct a projection matrix relating the properties of the red side (1280\,$<$$\lambda_{\rm rest}$\,$<$\,2900\,\textup{\AA}) of the quasars  to the Ly$\alpha$ emission line properties. They could predict quasars Ly$\alpha$ spectral region with the precision of 6$-$12\,\%. Moreover, \citet{2020MNRAS.493.4256A} used neural network to predict the blue-side PCA coefficients using the red-side PCA coefficients as the input to the model. Their predicted blue-side PCA coefficients could then be used to reconstruct the quasar continuum around the Ly$\alpha$ spectral region. The most sophisticated model found in the literature, is introduced by \citet{2017MNRAS.466.1814G}, which predicts the quasar Ly$\alpha$ emission with $\sim$\,9\,\% precision. Their approach was based on utilizing a co-variance matrix of emission line properties extracted through applying Gaussian fits to broad emission lines of Ly$\alpha$, Si\,{\sc iv}+O\,{\sc iv]}, C\,{\sc iv}, and C\,{\sc iii]}.

In this work, we develop a model based on deep neural network to predict quasars continua around Ly$\alpha$ spectral region. We aim at decoding the wealth of information hidden in the structure of the Si\,{\sc iv}, C\,{\sc iv}, and C\,{\sc iii]} emission lines in order to predict the shape and flux of the Ly$\alpha$ emission line.

\begin{figure}
\centering
\begin{tabular}{cc}
\includegraphics[width=0.99\hsize]{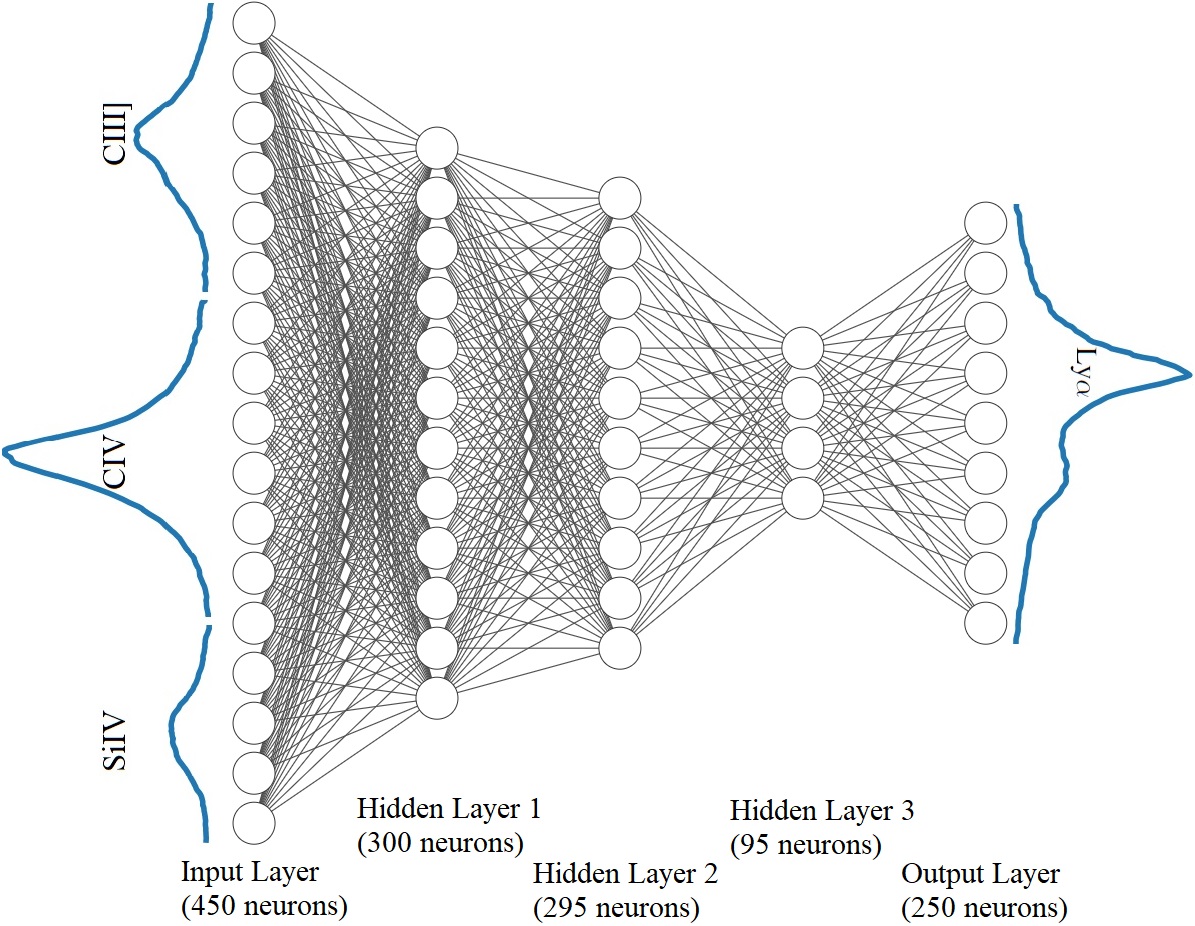}
\end{tabular}
\caption{A schematic showing the architecture of our best neural network model (see the text). The broad Si\,{\sc iv}, C\,{\sc iv}, and C\,{\sc iii]} emission lines are the input, and the output is the quasar continuum around the Ly$\alpha$ emission line.}
 \label{NNStructure}
\end{figure}

\section{deep learning method}

\subsection{Deep Neural Network Architecture}

The deep neural network used in this work is based on the feed-forward calculation and error back-propagation algorithms. The network has an input layer, hidden layers, and an output layer.  These layers are composed of different number of neurons. Figure\,\ref{NNStructure} shows an example of a neural network architecture. Fundamentally, a deep neural network is a function that maps input data to their associated output data through some data transformation that occurs in the hidden layers. The transformation implemented by a hidden layer is parameterized by the weights of the layer's neurons. 

Each neuron in the hidden layer needs to take the following two steps in order to produce an appropriate output which could be fed into the next layer. First, it should calculate the weighted sum of the output of all the neurons from the previous layer. For example, for the $j^{th}$ neuron in the $k^{th}$ layer, the following expression should be calculated

\begin{equation} \label{eq:1}
p_{j}^{k}(x_{1}, ..., x_{n}) = \sum_{i=1}^{n}\omega_{i,j}^{k}x_{i} + b_{j}^{k},
\end{equation}

\noindent
where $\omega_{i,j}$ and $b_{j}^{k}$ are the weights and biases of the layers that are determined through the error back-propagation technique, and $x_{i}$ are input from the previous layer. Second, the weighted sum from equation\,\ref{eq:1} should then be passed through an activation function. Activation functions add non-linearity to the model. Here, we employ the rectified linear unit, or the relu function, as the activator for the neurons in the hidden layers. The relu function is defined as 

\begin{align}  \label{eq:2}
R(x) = \left\{ \begin{array}{cc}
                    0 & \hspace{5mm} x < 0 \\
                   x  & \hspace{5mm} x \ge 0 \\
                   \end{array} \right.
\end{align}

\noindent
In other words, each neuron in the hidden layers takes the output of all the neurons from the previous layer and returns the following expression

\begin{equation} \label{eq:3}
R(p_{j}^{k}(x_{1}, ..., x_{n})) = max(0, \sum_{i=1}^{n}\omega_{i,j}^{k}x_{i} + b_{j}^{k})
\end{equation}

\begin{figure*}
\centering
\begin{tabular}{c c}
\includegraphics[width=0.49\hsize]{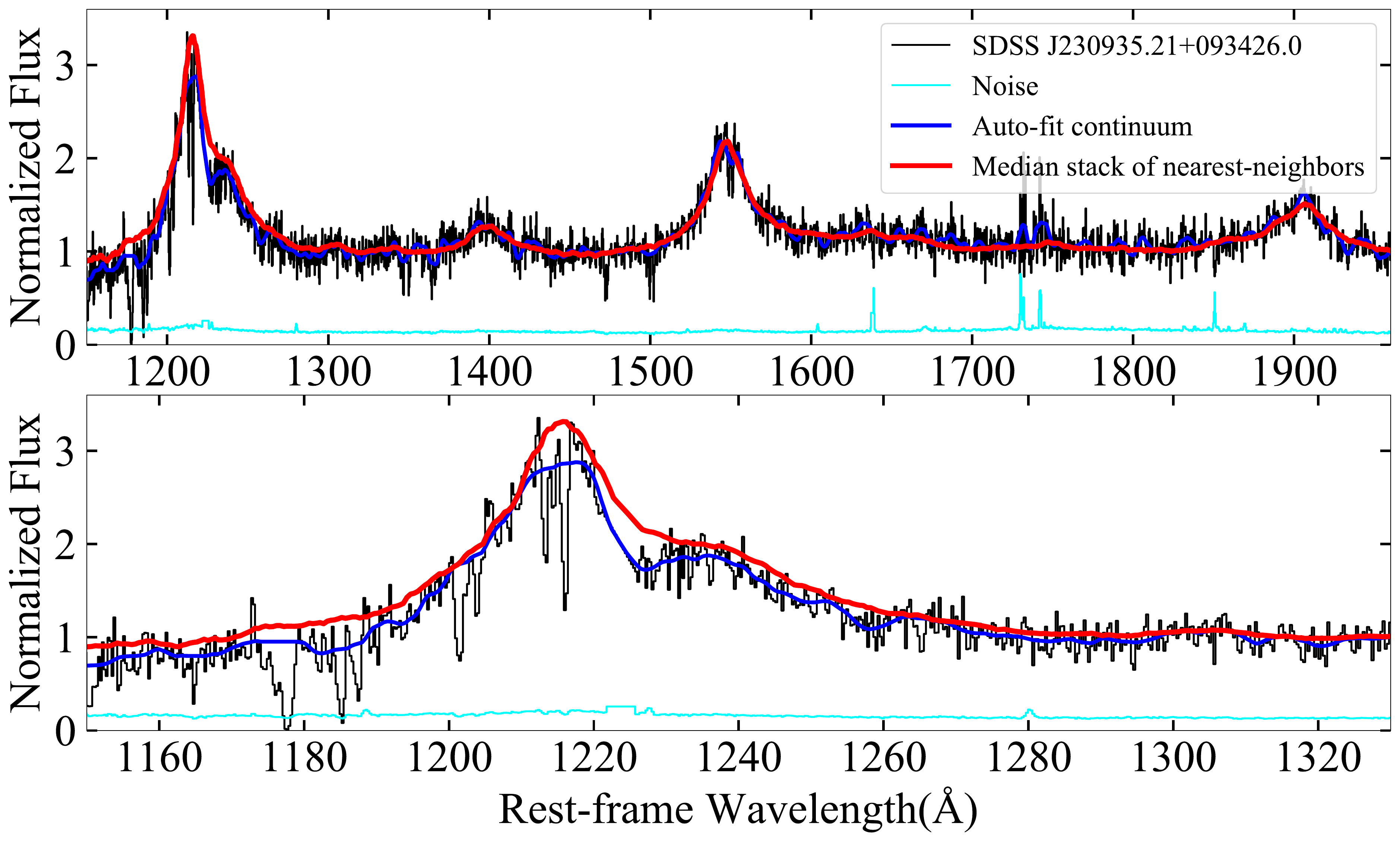} 
\includegraphics[width=0.49\hsize]{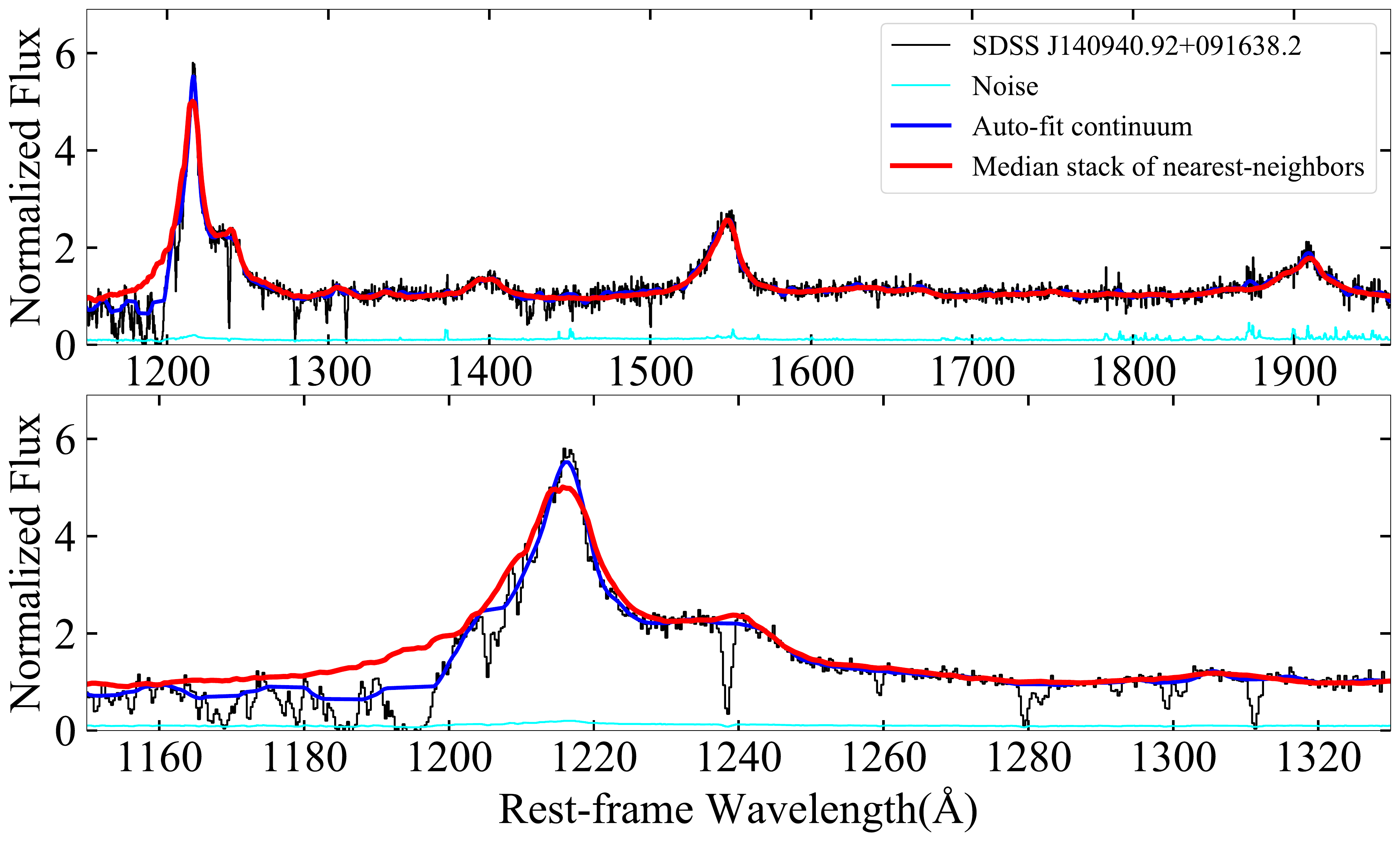} \\
\includegraphics[width=0.49\hsize]{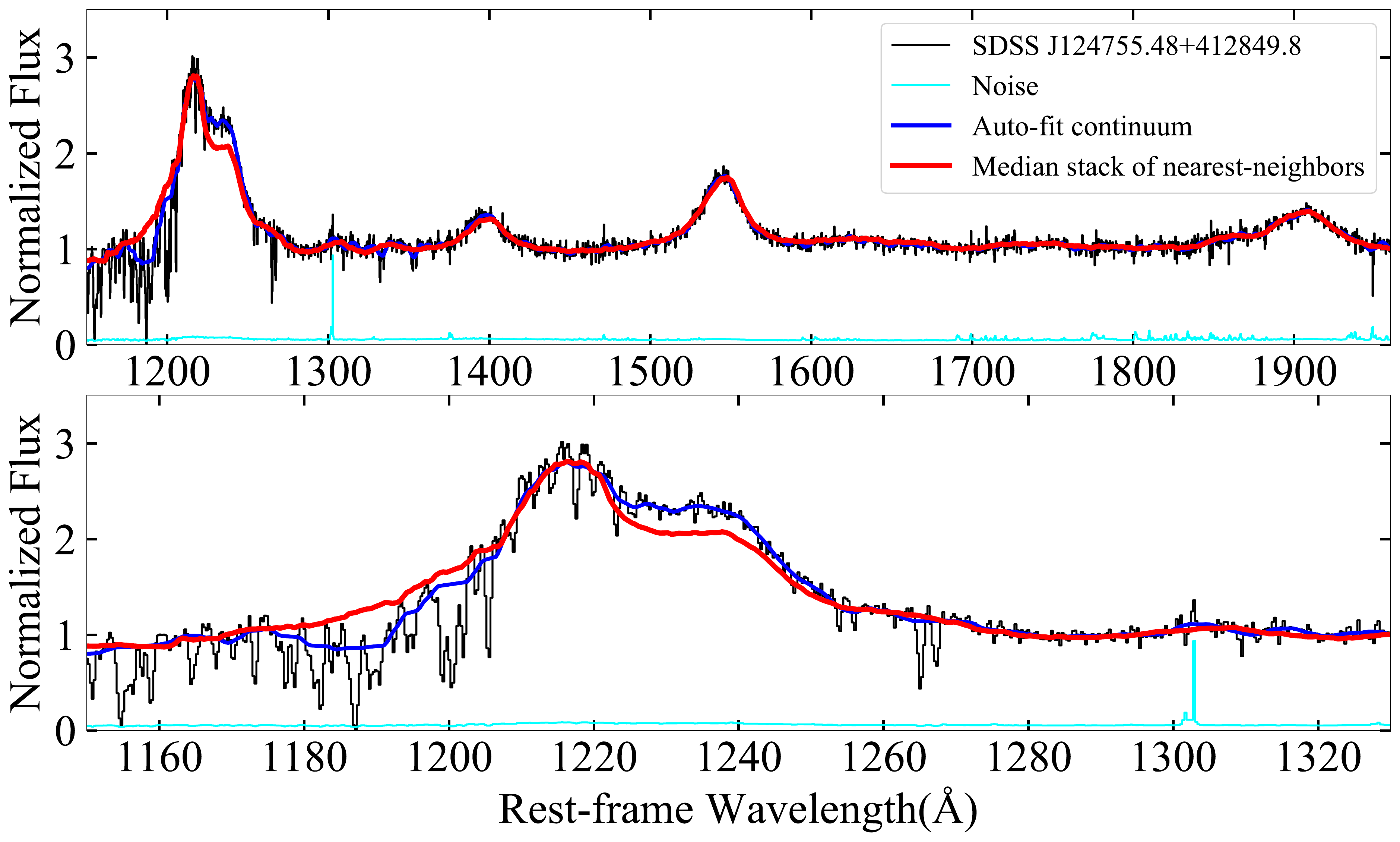} 
\includegraphics[width=0.49\hsize]{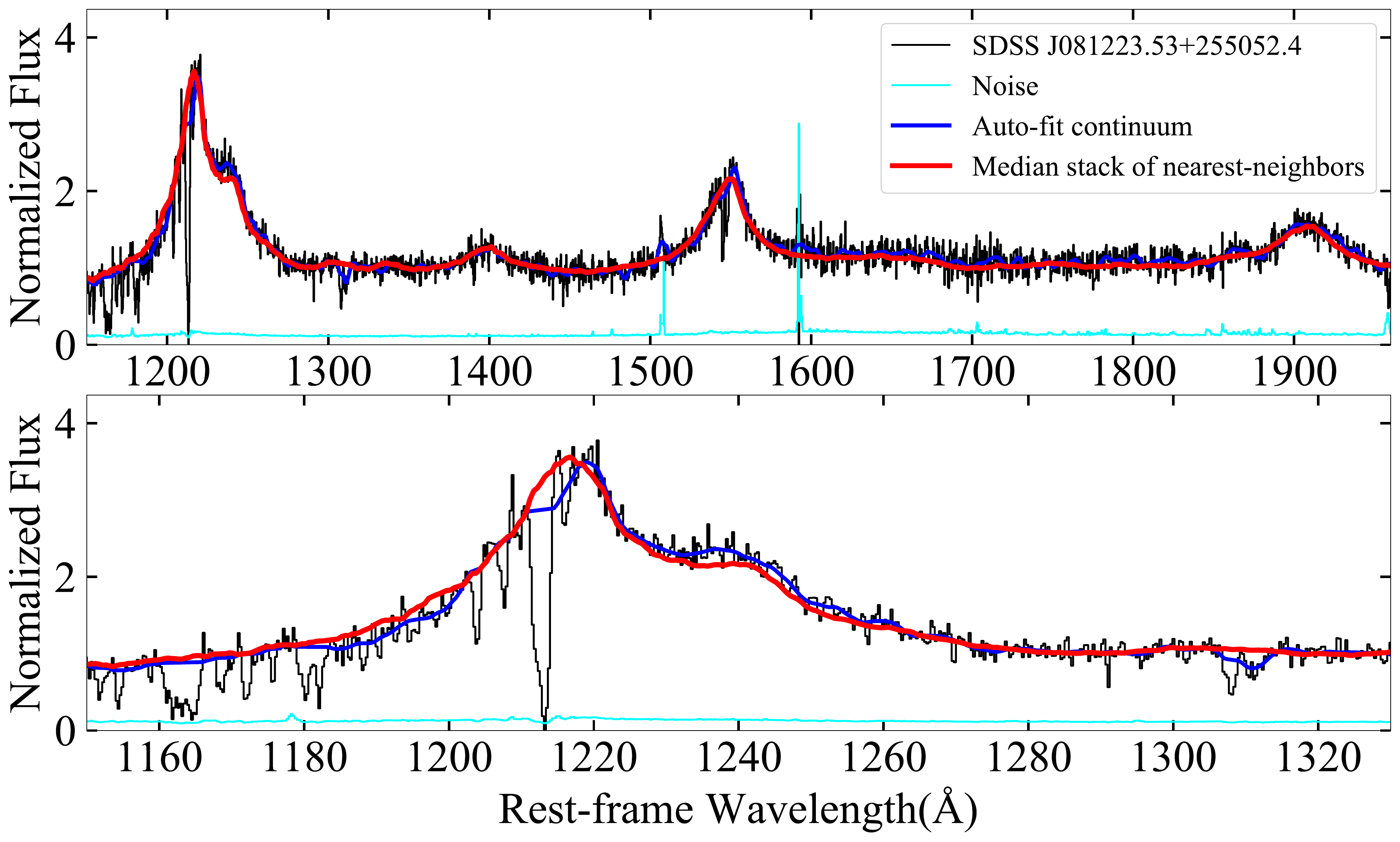} 
\end{tabular}
\caption{Four randomly selected SDSS-DR14 quasar spectra. In each panel, the blue curve shows the auto-fit performed by the Savitzky-Golay filtering, and the red curve shows the median stack of nearest neighbor quasars. The error spectrum is shown in cyan.}
 \label{autofit}
\end{figure*}

\subsection{Data}

We use quasar spectra from the SDSS-DR14 to train our neural network and evaluate its performance. To make sure that the quasar Ly$\alpha$ and C\,{\sc iii]} emission lines always fall on the observed spectral window, we only take into account quasars with emission redshift from $z_{\rm QSO}$\,$=$\,2.0 to 4.3. Quasars with broad absorption lines (BAL) are excluded from the sample. In the SDSS-DR14 catalog of \citet{2018A&A...613A..51P}, quasars with BALs are recognized by their nonzero BAL\_FLAG\_VI index. We also exclude quasar spectra with the signal-to-noise ratio (SNR) below 15. The SNR is measured over the spectral region between 1620 to 1740\,\textup{\AA} in the quasar rest frame. Applying these constraints on the SDSS-DR14 spectra returns 48\,703 quasars. Each quasar spectrum is then shifted to the quasar rest-frame, rebinned to the same wavelength grid (with the pixel size of 0.5\,\textup{\AA}),  and normalized by fitting a power law function on the regions outside the emission lines. 

\begin{figure}
\centering
\begin{tabular}{c}
\includegraphics[width=0.99\hsize]{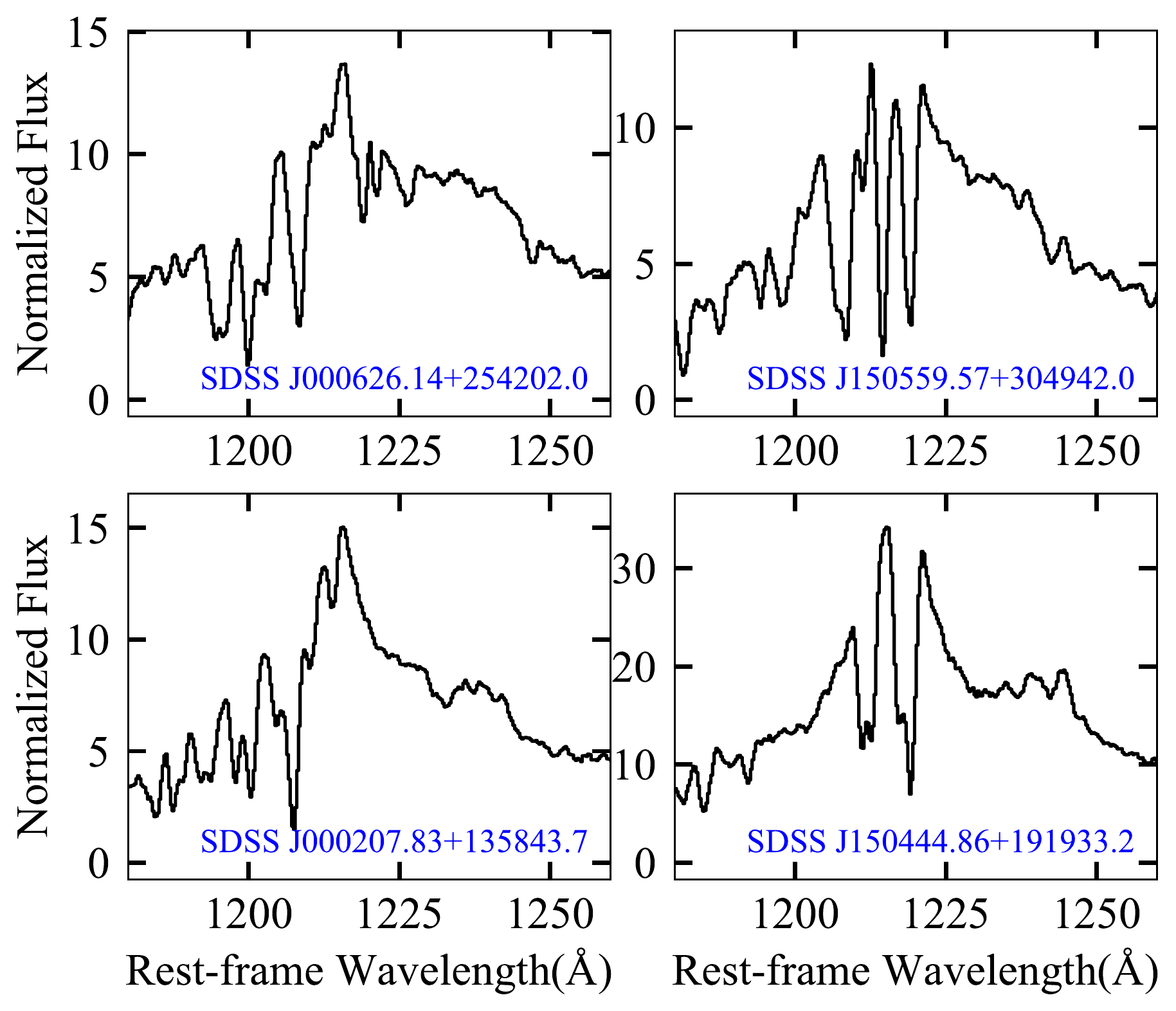}
\end{tabular}
\caption{Four examples of quasar spectra exhibiting numerous strong absorption lines in the Ly$\alpha$ spectral region. Quasar spectra with such absorption features are excluded from our final sample.}
 \label{rejqsos}
\end{figure}

We then automatically fit the whole spectrum (i.e. including both the continuum and emission lines) by iteratively applying the Savitzky-Golay filtering \citep{1964AnaCh..36.1627S} and removing pixels that deviate more than two standard deviations from the fit. For each pixel, the associated standard deviation is measured locally by taking into account regions close to that pixel. Figure\,\ref{autofit} shows four randomly selected quasar spectra with the auto-fit continua over-plotted as the blue curves. The presence of strong and/or numerous adjacent absorption lines in the Ly$\alpha$ spectral region could in principle significantly change the shape of a broad Ly$\alpha$ emission line. Therefore, in these cases, the auto-fit continuum could not be a good representative of the unabsorbed Ly$\alpha$ emission. To minimize this effect, we exclude quasar spectra that have such contaminating absorption features around their Ly$\alpha$ emission line. Figure\,\ref{rejqsos} shows four examples of quasar spectra that were rejected from our final sample, mainly due to the presence of numerous absorption features around the Ly$\alpha$ spectral region.

By visually inspecting all 48\,703 spectra and excluding those with unwanted features, we are left with 17\,870 spectra.
This is our final quasars sample. Although quasars with strong associated absorption in the Ly$\alpha$ region are visually excluded, the presence of narrower absorption, in some cases, could still affect the auto-fit continuum, albeit less severely (see Fig.\,\ref{autofit}). To further clean up the sample, we follow \citet{2018ApJ...864..143D} and replace each spectrum with a median stack of its 20 nearest-neighbors in the sample. The auto-fit continua of quasars are used in the nearest-neighbor analysis. We use the Euclidean distance

\begin{equation} \label{eq:3}
D_{ij} = \sqrt{\sum_{\lambda} (C_{\lambda,i} - C_{\lambda,j})},
\end{equation}

\noindent
to find the neighboring quasars. Here, $C_{\lambda,i}$ and $C_{\lambda,j}$ are the normalized auto-fit continua of two different quasars. Since a quasar's broad Ly$\alpha$ emission (especially its blue wing) is almost always contaminated by narrow absorption from the Ly$\alpha$ forest, the nearest-neighbors are defined only using pixels with $\lambda$\,$>$\,1280\,\textup{\AA}. This helps us to avoid combining spectra with similar absorption occurring in the Ly$\alpha$ region. In Fig.\,\ref{autofit}, the nearest-neighbor stack spectra are shown as red curves.

In our neural network models, we use the Si\,{\sc iv}, C\,{\sc iv}, and C\,{\sc iii]} emission lines as the input data, and predict the Ly$\alpha$ emission line as the output. More specifically, we only take into account the flux values of the pixels corresponding to these emission lines. The Si\,{\sc iv}, C\,{\sc iv}, and C\,{\sc iii]} emission lines each extends over 120 (1370$-$1430\,\textup{\AA}), 170 (1505$-$1590\,\textup{\AA}), and 160 (1865$-$1945\,\textup{\AA}) spectral pixels, respectively. Therefore, the number of neurons in the input layer is 450, which equals to the total number of spectral pixels of these emission lines. Similarly, the number of neurons in the output layer is 240, which is equal to the number of spectral pixels in the Ly$\alpha$ emission line region from 1170 to 1290\,\textup{\AA}.

Since our neural network models are based on mapping the spectral pixels of the Si\,{\sc iv}, C\,{\sc iv}, and C\,{\sc iii]} emission lines onto  those of the Ly$\alpha$ emission line, it is very important that before constructing any model, these emission lines are well aligned with each other in individual spectra. Since inaccurate redshift of a quasar could in principle lead to the misalignment of these emission lines, we therefore try to readjust the redshift of each quasar in our final sample. To do this, we first create a template spectrum by median-stacking all 17\,870 quasar spectra from the final sample. We then align the auto-fit continuum of each quasar with the template spectrum in such a way that the emission lines in the two spectra are well aligned. These readjusted quasar spectra are then used in training and testing our models.

\subsection{Train, validation, and test data set split}

We split our sample of 17\,870 quasars into three sub-samples, namely, the training , validation, and test data, with each sub-sample containing 10722 (60\,\%), 3574 (20\,\%), and 3574 (20\,\%) data points, respectively. Generally, in machine learning, the models are trained on the training data and evaluated using the validation data. However, although the validation data is not used to train the models (i.e. learning the weights and biases), it still indirectly influences the fine-tuning of the model parameters. This is because we select the best model as a model with the best performance on the validation data. Therefore, the evaluation of the models becomes biased.

To circumvent this problem, the test data is used to provide an unbiased evaluation of the model. We use the test data only after we have chosen the final model. We recall that the best final model is the one with the best performance on the validation data. Applying our final model on the test data allows us to see how our preferred model is going to perform on completely unseen data.

\begin{figure*}
\centering
\begin{tabular}{c}
\includegraphics[width=0.99\hsize]{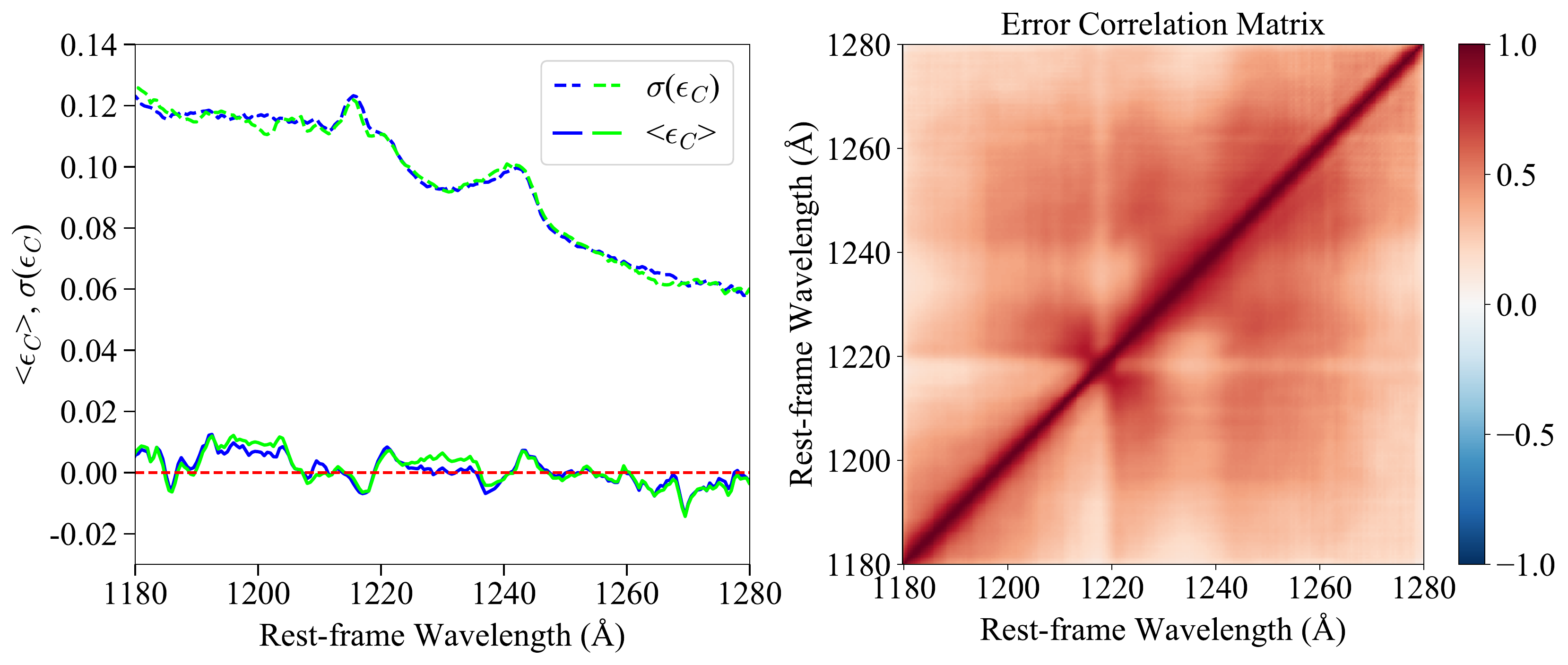}
\end{tabular}
\caption{Left: relative uncertainty (1\,$\sigma$ error, dashed lines) and mean bias (solid lines) of the neural network prediction of the Ly$\alpha$ spectral region for the validation (blue lines) and test (green lines) data. Right: correlation matrix of our neural network prediction errors.}
 \label{resultx}
\end{figure*}

\subsection{Hyper-parameters Tuning}

We train our neural network models in Python\,3.6 environment using the open source libraries Keras and Tensorflow. Since we are dealing with a regression problem, \textit{mean-squared-error} is chosen as the loss function. The loss function is minimized using the \textit{Adam} optimizer. The training epoch is set to 1000, and the early stopping of training is activated via the \textit{EarlyStopping} callback in Keras. The activation of this callback prevents the overfitting of the training data set by stopping the training process once the model is no more getting improved.

One of the most challenging parts of training a neural network model is the tuning of its hyper-parameters, especially the number of hidden layers and their associated neurons which control the architecture or topology of the network. Although the number of neurons in the input and output layers are uniquely determined by the data itself, the number of hidden layers and their neurons are left as free parameters and need to be optimally determined. We use random search to find these parameters. Compared to other searching strategies (e.g. trial and error and grid search), random search allows us to explore the hyper-parameter space more efficiently. This in turn helps us to find the best configuration in fewer iterations.

To find the model with the optimal configuration, we construct 50\,000 neural networks with random number of hidden layers and neurons. The number of hidden layers are randomly chosen to be between 1 and 4, and the number of neurons comprising each of these hidden layers is found through randomly generating multiple of 5 numbers between 10 and 300. In the next section, we define a metric which helps us to single out the best model out of all randomly generated models.

\begin{figure*}
\centering
\begin{tabular}{cc}
\includegraphics[width=0.49\hsize]{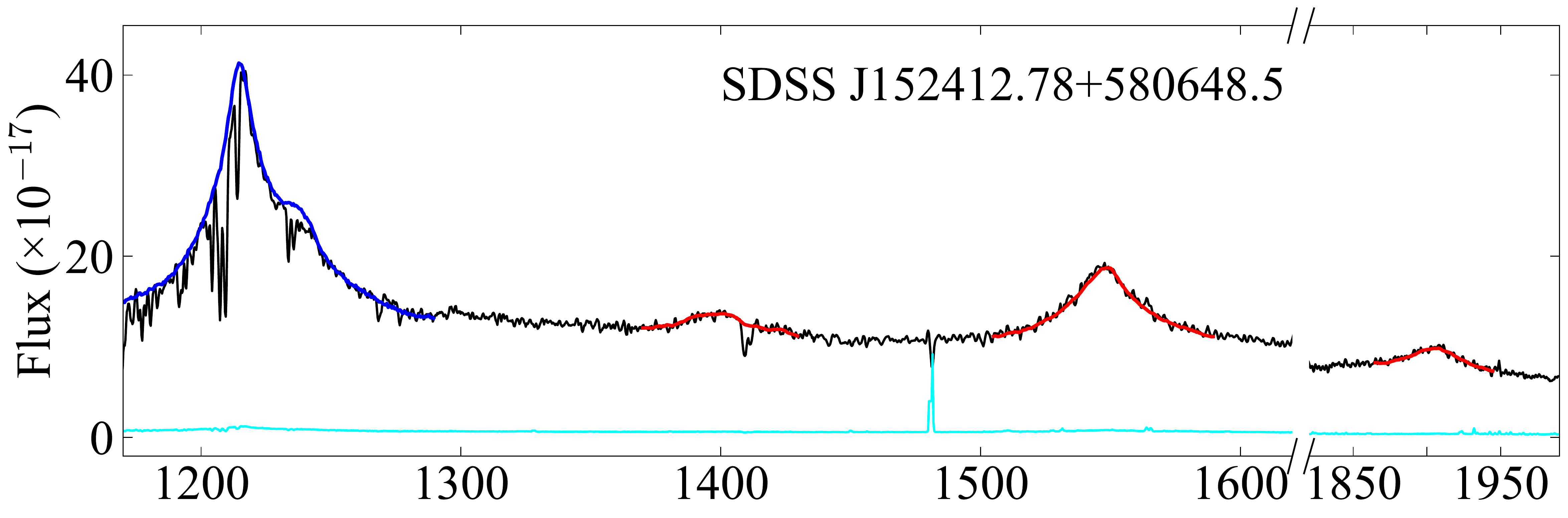}
\includegraphics[width=0.49\hsize]{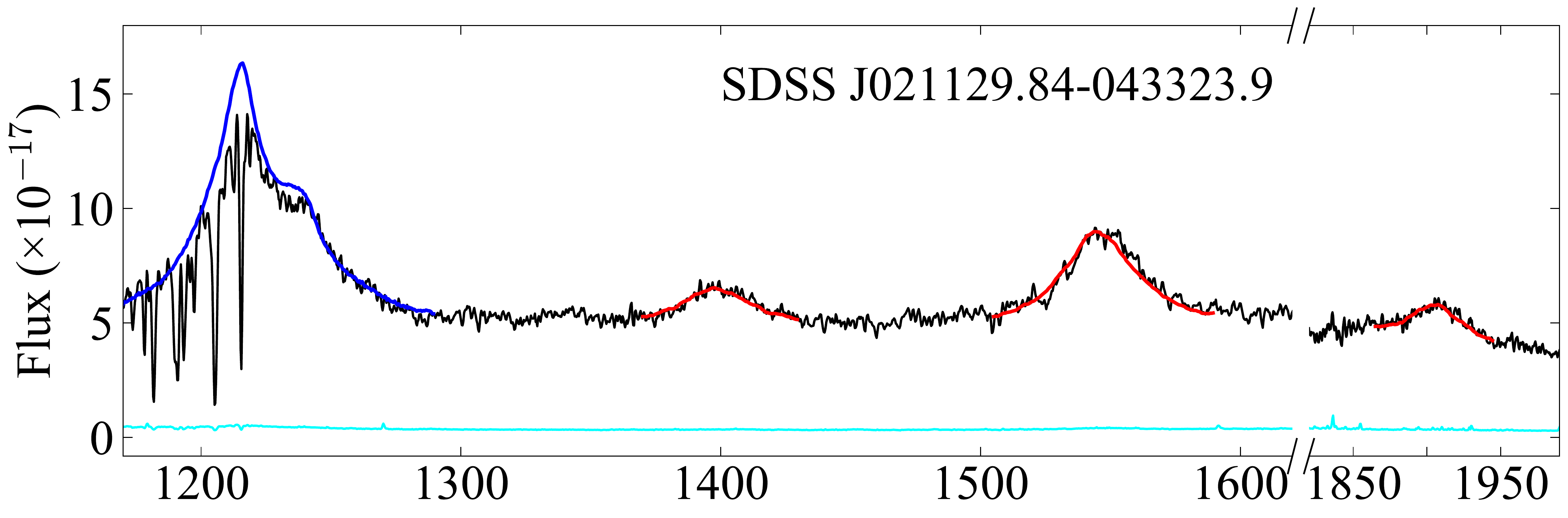} \\
\includegraphics[width=0.49\hsize]{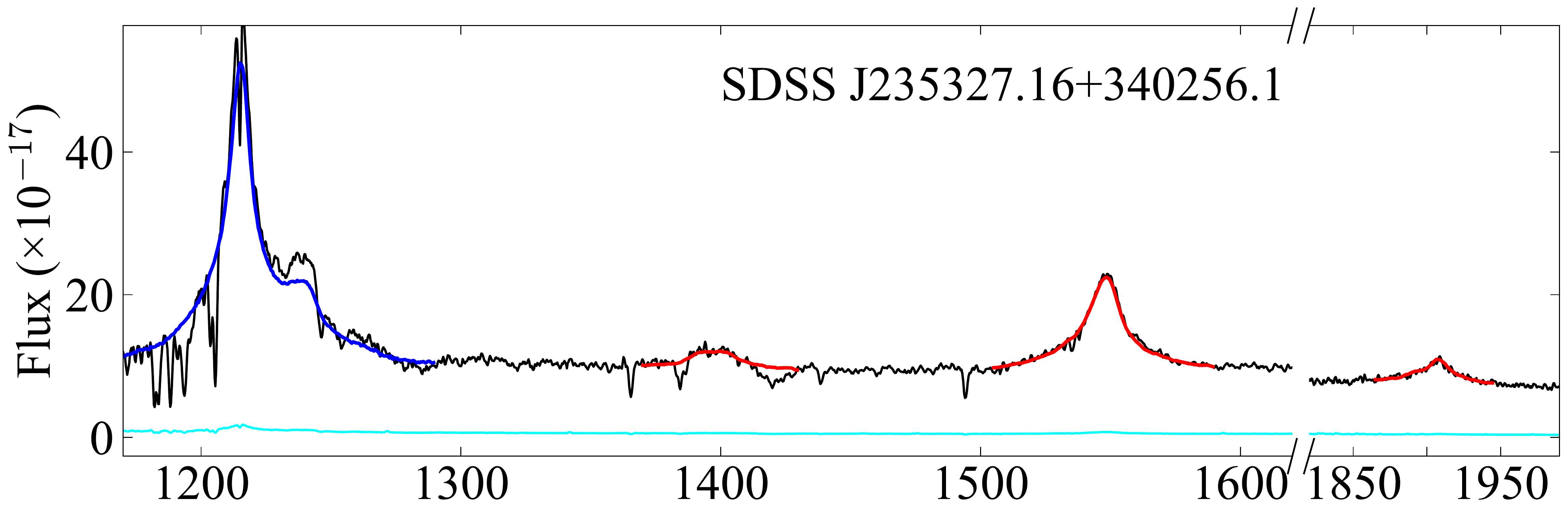}
\includegraphics[width=0.49\hsize]{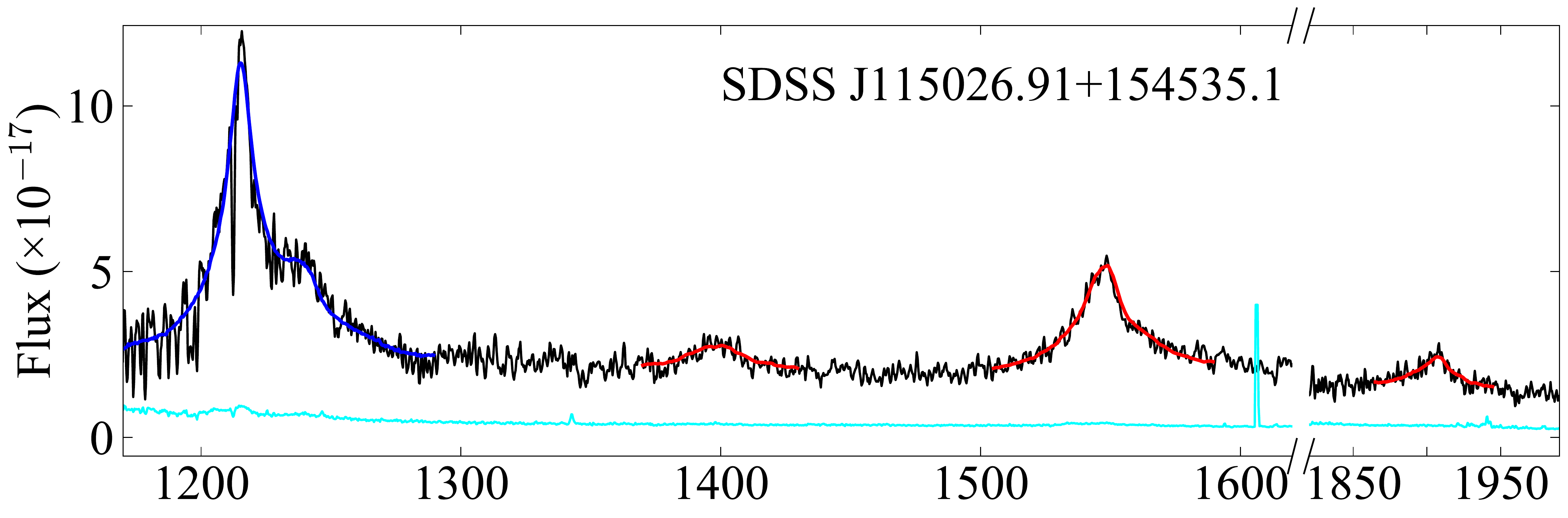} \\
\includegraphics[width=0.49\hsize]{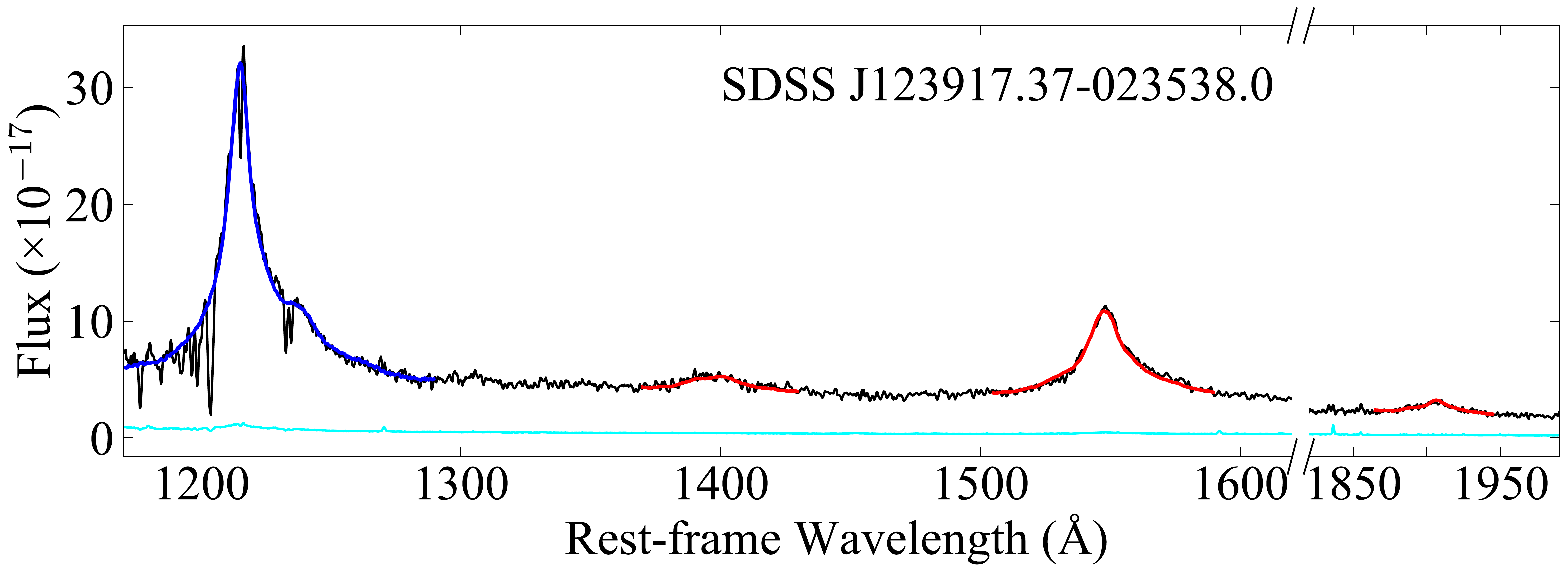}
\includegraphics[width=0.49\hsize]{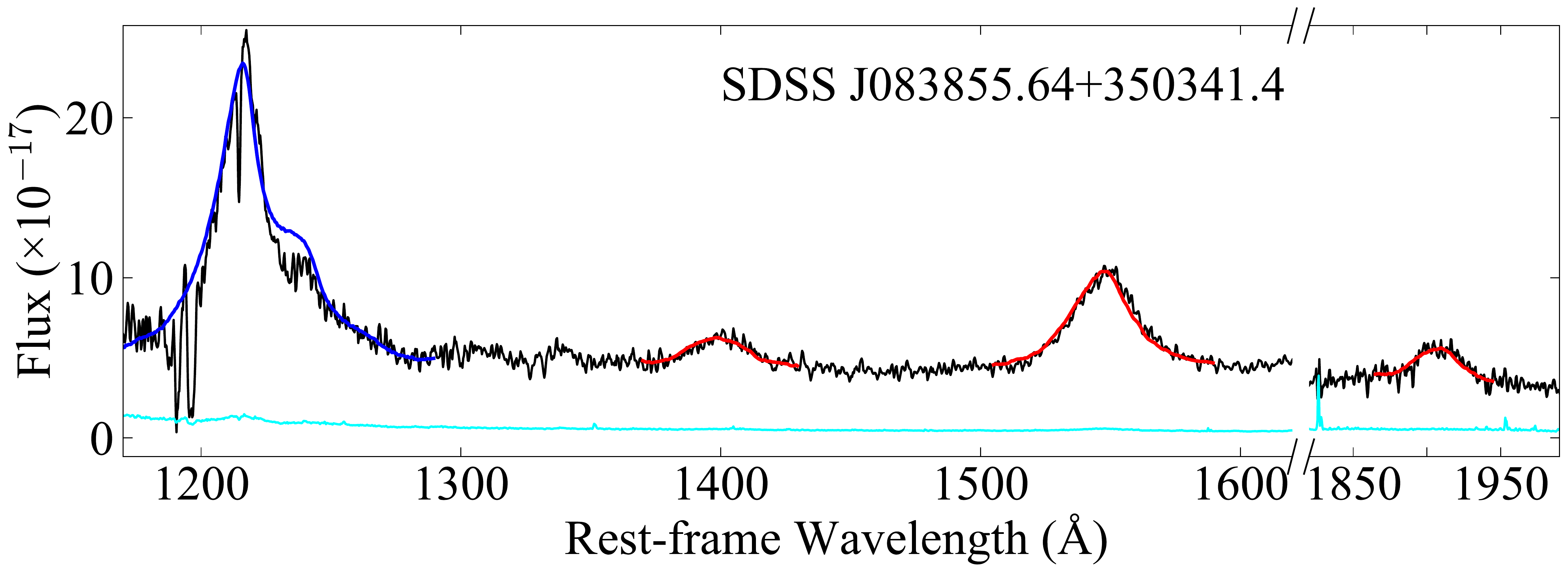}
\end{tabular}
\caption{Neural network predictions of the Ly$\alpha$ emission in six randomly selected quasars from the test data set. In each panel, the blue curve is the prediction and the red curves are the input to the model. The error spectrum is shown in cyan, and the y-axis is in erg\,s$^{-1}$\,cm$^{-2}$\,\textup{\AA}$^{-1}$.}
 \label{prediction}
\end{figure*}

\section{Results}

Although the loss function allows us to monitor the convergence of each model, this is not an appropriate metric to evaluate the predictive power of our neural networks. Therefore, we use another metric to asses the performance of the models. Following \citet{2018ApJ...864..143D}, for each model, we calculate the relative continuum error as follows

\begin{equation} \label{eq:4}
\epsilon_{C}= \frac{|F_{pred} - F_{true}|}{F_{true}}
\end{equation}

\noindent
where $F_{pred}$ is the flux predicted by the model and $F_{true}$ is the \emph{true} flux.  Here, we consider the Savitzky-Golay auto-fit continuum of each quasar as the true flux.

The standard deviation of $\epsilon_{C}$ is used as a metric to find the best model out of our randomly generated models.  Checking $\epsilon_{C}$ for all the models, we could find a model which exhibits the smallest error across the wavelength range relevant to damping wing studies, i.e. 1210\,$<$\,$\lambda$\,$<$\,1250\,\textup{\AA}. We chose this model as our best neural network model. This model has three hidden layers. The number of neurons in the first, second and third hidden layers is 300, 295, and 95, respectively (see Fig.\,\ref{NNStructure}). We show in Figure\,\ref{resultx} the mean (solid lines) and standard deviation (dashed lines) of $\epsilon_{C}$ as a function of rest frame wavelength for our best model. In this figure, the blue (resp. green) curves show the results on the validation (resp. test) data. Similar to \citet{2018ApJ...864..143D}, at each wavelength in Fig.\,\ref{resultx}, individual pixels that deviate more than three standard deviations from the mean were clipped. After three iterations of clipping, $\sim$\,1\% of the pixels in the Ly$\alpha$ spectral region were masked.

As shown in Fig.\,\ref{resultx}, the error and the bias for both validation and test data follow each other very closely. This implies that our final model is very good in generalizing and that it would perform well on unseen data.  The 1\,$\sigma$ error in the damping wing regions is $\sim$\,6$-$12\,\% with a bias of $\lesssim$\,1\%, which is comparable to the $\sim$\,6$-$12\,\% error in \citet{2018ApJ...864..143D} and $\sim$\,9\% error in \citet{2017MNRAS.466.1814G}. We also show in Fig.\,\ref{resultx} (right panel) the correlation matrix of the errors in the continuum prediction, i.e. $\epsilon_{C}$. The plot shows strong correlation across neighboring pixels. This correlation is partly due to the correlated errors in the autofit continuum. Figure\,\ref{prediction} shows examples of our neural network predictions of the Ly$\alpha$ emission in six randomly selected quasars from the test data set. Here, in each panel, the blue curve is the prediction and the red curves are the input to the model. Our model is available on GitHub.\footnote{\url{https://github.com/hfathie/qso/blob/master/NN_Model.h5}}

\section{Summary and Conclusion}

In this work, we have developed a deep neural network model to predict quasars continua in the Ly$\alpha$ spectral region. The input to the network is the normalized flux of the individual pixels corresponding to Si\,{\sc iv}, C\,{\sc iv}, and C\,{\sc iii]} emission lines. The model then returns the predicted broad Ly$\alpha$ emission line. We trained, tested and evaluated the performance of the model using 17\,870 quasar spectra from the SDSS-DR14. By testing the model performance on the test data, we found that the model can predict the quasar Ly$\alpha$ emission line of an individual quasar to $\sim$\,6$-$12\,\% precision with a very small bias ($\lesssim$\,1\,\%). We can use this model to estimate the H\,{\sc i} column density of eclipsing and ghostly DLAs as the presence of the strong Ly$\alpha$ absorption in these systems severely contaminates the flux and the shape of the quasar continuum around Ly$\alpha$ spectral region. Moreover, this work could also be used to study the state of the intergalactic medium during the epoch of reionization by reconstructing the damping wing of the Ly$\alpha$ absorption in the spectra of high redshift quasars.

\section*{Acknowledgements}
The author would like to thank the referee for the constructive comments which improved the quality of the paper.
\noindent
Funding for SDSS-III has been provided by the Alfred P. Sloan Foundation, the Participating Institutions, the National Science Foundation, and the U.S. Department of Energy Office of Science. The SDSS-III web site is \url{http://www.sdss3.org/}.
\noindent
SDSS-III is managed by the Astrophysical Research Consortium for the Participating Institutions of the SDSS-III Collaboration including the University of Arizona, the Brazilian Participation Group, Brookhaven National Laboratory, Carnegie Mellon University, University of Florida, the French Participation Group, the German Participation Group, Harvard University, the Instituto de Astrofisica de Canarias, the Michigan State/Notre Dame/JINA Participation Group, Johns Hopkins University, Lawrence Berkeley National Laboratory, Max Planck Institute for Astrophysics, Max Planck Institute for Extraterrestrial Physics, New Mexico State University, New York University, Ohio State University, Pennsylvania State University, University of Portsmouth, Princeton University, the Spanish Participation Group, University of Tokyo, University of Utah, Vanderbilt University, University of Virginia, University of Washington, and Yale University.

\bibliographystyle{aasjournal}
\bibliography{ref}

\end{document}